\documentclass[11pt]{article}
\usepackage[totalwidth=387pt,totalheight=590pt]{geometry}
\usepackage{amsmath,amssymb,epsf,graphics,graphicx,mathrsfs,eufrak}
\usepackage{enumerate}
\usepackage{psfrag}
\newcommand{\secref}[1]{\S\ref{#1}}
\newcommand{\bm}[1]{{\bf{#1}}}
\newcommand{\de}{\partial}
\newcommand{\la}{\lambda}
\newcommand{\bz}{\bar{z}}
\newcommand{\om}{\omega}

\newtheorem{conjecture}{Conjecture}[section]
\newcommand{\physics}[1]{{[arXiv:physics/#1]}}

\newcommand{\quantph}[1]{{[arXiv:quant-ph/#1]}}

\newcommand{\ZZ}{\mathbb{Z}}

\newcommand\blank[1]{}

\newcommand\toline[1]{--#1}

\renewcommand{\hat}{\widehat}
\newcommand\eq{\begin{equation}}
\newcommand\en{\end{equation}}
\newcommand\bea{\begin{eqnarray}}
\newcommand\eea{\end{eqnarray}}

\newcommand\ba{\(\begin{array}}
\newcommand\ea{\end{array}\)}

\newcommand\al{\alpha}

\newcommand\te{\theta}

\renewcommand\al{M}
\begin{document}
\begin{titlepage}
\vskip 0.5cm
\vskip 1.2cm
\begin{center}
{\Large{\bf The Bethe Ansatz and  the Tzitz\'eica-Bullough-Dodd
equation}}
\end{center}
\vskip 0.8cm \centerline{Patrick Dorey$^{1}$, Simone Faldella$^{2}$,
Stefano Negro$^3$ and Roberto Tateo$^3$} \vskip 0.9cm

\vskip 0.3cm \centerline{${}^{1}$\sl\small Department of Mathematical
Sciences, Durham University,} \centerline{\sl\small South Road, Durham
DH1 3LE, United Kingdom}

\vskip 0.3cm \centerline{${}^2$\sl\small Institut de Math\'ematiques
de Bourgogne, Universit\'e de Bourgogne,} \centerline{\sl\small  9 av.
Alain Savary - B.P. 47 870, 21078 Dijon, France}
\vskip 0.3cm \centerline{${}^{3}$\sl\small Dip.\ di Fisica 
and INFN, Universit\`a di Torino,} \centerline{\sl\small Via P.\
Giuria 1, 10125 Torino, Italy}
\vskip 0.2cm \centerline{E-mails:}
\centerline{p.e.dorey@durham.ac.uk,  simone.faldella@u-bourgogne.fr,}
\centerline{negro@to.infn.it, tateo@to.infn.it}

\vskip 1.25cm
\begin{abstract}
\noindent
The theory of classically integrable nonlinear wave equations, and the
Bethe Ansatz  systems describing massive quantum field theories
defined on an infinite cylinder, are related by an important
mathematical correspondence that still lacks a satisfactory
physical interpretation.  In this paper we shall describe this link
for the case of the classical and  quantum versions of the
(Tzitz\'eica-)Bullough-Dodd model.   

\end{abstract}

\bigskip

{\small


}

\end{titlepage}
\setcounter{footnote}{0}
\def\thefootnote{\fnsymbol{footnote}}
%
\section{Introduction}
An interesting correspondence was discovered in \cite{Dorey:1998pt}
between the functional approach to the spectral theory of  ordinary
differential equations, mainly developed by  Sibuya, Voros  and
collaborators \cite{Sha, Voros}, and a series of works  by
Bazhanov, Lukyanov and Zamolodchikov \cite{Bazhanov:1994ft,
Bazhanov:1996dr,Bazhanov:1998dq}, where  earlier results of Baxter
\cite{Bax} were extended to the conformal field theories 
governing  the  continuum limits of certain integrable models (IM) 
on a 2d lattice. 

It turns out that the initial observation of \cite{Dorey:1998pt}
was the first hint of a very general mathematical scheme  --
an `ODE/IM correspondence' --  involving wide classes of multi-parameter 
ordinary   differential  equations.  Their generalized
eigenvalue problems turn out to be
constrained by  the same Bethe Ansatz  equations as arise in
the conformal field theory 
limits of certain integrable vertex  models, related to Lie algebras
\cite{Bazhanov:1998wj,Suzuki:1999rj,
Suzuki:1999hu, Dorey:1999pv,  Dorey:2006an, Dorey:2007zx, Dorey:2009xa}.   

A precise relationship between elements of this new scheme  
and the ${\cal P}{\cal T}$-symmetric quantum mechanical
models of Bender, Boettcher and Meisinger 
\cite{BB,BBN} was established in \cite{Dorey:1999uk}. This has
led to proofs of spectral reality
in certain cases \cite{Dorey:2001uw}, that were further
generalised in \cite{Shin:2002vu},
and insights into the loss of
reality in others~\cite{Dorey:2009tc}.
Applications of the correspondence are also being found in various
problems arising in condensed matter physics
\cite{Bazhanov:2003ua,Gritsev:2006,Lamacraft:2008}.

A generalization of this correspondence to off-critical, or massive,
quantum field theories 
was subsequently  obtained by Gaiotto, Moore and
Neitzke \cite{Gaiotto:2009hg} following a different
chain of discoveries, relating to string and gauge theories.
As a net outcome of a surprising series of
mathematical connections, also bringing in the AdS/CFT correspondence,
gluon scattering amplitudes in ${\cal N}=4$
Super Yang-Mills can be now studied using powerful tools from the
theory of integrable models~\cite{Alday:2009dv}.

A further important step for the understanding of the relationship
between the AdS/CFT-related results and the original formulation of
the ODE/IM correspondence was made  by Lukyanov and Zamolodchikov  in
\cite{Lukyanov:2010rn}.  They showed how to link the classical
sinh-Gordon equation to the quantum massive sine(h)-Gordon model,
generalizing the earlier works on the ODE/IM correspondence. In doing
this, they brought into play all the formalism developed in more
than fifty years of soliton theory and the study of exactly-solvable 
nonlinear partial
differential equations (of which sinh-Gordon is an exemplar), in
particular the role played by the association between a nonlinear
classical field theory and certain sets of linear differential
equations.

In this paper we shall discuss another example of an off-critical
generalization of the ODE/IM correspondence, which
will be
built starting from the definition of a particular model on the
classical side. This model is often referred to as the
Bullough-Dodd model \cite{Dodd:1977bi}, though the special properties
of the corresponding partial differential equation were noticed as
long ago as 1907, by Tzitz\'eica \cite{Tzitzeica}.
Like the sinh-Gordon model, it is an example
of an affine Toda field theory, or 2d Toda chain, as will be explained
in \secref{BDdef-sec}. 
The quantum version of the Bullough-Dodd   model, also known as   the
Izergin-Korepin model \cite{IzKo},  plays an important role in
framework of 
massive 2D quantum field theories corresponding to integrable
perturbations of the minimal series of conformal field theories
\cite{ZamInt}. 
Together with its purely-imaginary  coupling  version  
it has been directly studied using the exact S-matrix approach
\cite{Smirnov:1991uw}.  
It is also related   to the  scaling limit of important 
statistical-mechanical  systems  
such as  the q-state Potts models \cite{Potts} and
the dilute $A_n$ 
models \cite{Warnaar:1992gj}.   

The analysis in this paper follows the steps taken in
\cite{Dorey:1999pv} and \cite{Lukyanov:2010rn}. 
Some further details
of the Bullough-Dodd model and the ODE/IM correspondence can be found
in \cite{TesiFaldella}, while a more comprehensive 
presentation encompassing the extension to other
affine Toda field theories  will be the subject of a forthcoming
publication. 
\section{The Bullough-Dodd model}
\label{BDdef-sec}
The one dimensional Toda chain is defined as:
\begin{equation}\label{1dToda}
 \partial^2_{t} 
\eta^k =2 e^{(2\eta^{k+1}-2\eta^{k})}-2 e^{(2\eta^k-2\eta^{k-1})},
\end{equation}
where $k$ enumerates and labels the fields.
This definition can be simply modified  to yield the $(1+1)$-dimensional
variant of \eqref{1dToda} \cite{Toda}:
\begin{equation}\label{2dToda}
 \partial^2_t \eta^k- \partial^2_x \eta^k = 2 e^{(2\eta^{k+1}-2\eta^{k})}-2 e^{(2\eta^k-2\eta^{k-1})}.
\end{equation}
Varying the periodicity conditions and symmetries of the fields
$\eta^k$, the system (\ref{2dToda}) leads to a whole family of non linear
equations which are solvable by means of the Inverse Scattering
Transform (see for example \cite{Ablowitz}). This involves
an associated linear system, which has to reproduce the nonlinear chain
through a compatibility condition. It turns out that the correct
choice for (\ref{2dToda}) is \cite{Toda}\,:
\begin{equation}
\left \{
\begin{aligned}\label{Toda-sys}
  X\Psi& \equiv (\de_t+V- i \la C_1- i \la^{-1}C_2)\Psi&=\,0\\
  T\Psi& \equiv (\de_x+W+ i \la C_1- i \la^{-1}C_2)\Psi&=\,0
 \end{aligned}\right.
\end{equation}
where $\la$ is the \textit{spectral parameter},
\begin{equation}
\begin{aligned}
 V_{ij}=\de_t\eta_i\delta_{ij}\,;~ &W_{ij}=\de_x\eta_i\delta_{ij}\,;~
C_{1ij}=C_{2ji}=C_j\delta_{i-1,j}\,,\\
&C_i=e^{(\eta^{i+1}-\eta^i)}
\end{aligned}
\end{equation}
and $\delta_{ij}$ is the Kronecker symbol which, in a case of a closed
chain consisting of $N$ elements, is defined as:
\begin{equation}
 \delta_{ij}=\left\{
\begin{aligned}
 1, \quad &\text{when} \quad i\equiv j \; \mod N; \\
 0, \quad &\text{in all other cases.}
\end{aligned}\right.
\end{equation}

It is also possible to give a Lagrangian formulation for
the system \eqref{2dToda}:
\begin{equation}
 \mathcal{L}=\sum_{k=1}^{N} \left(\frac1{2} \partial_{\mu} 
\eta^{k} \partial^{\mu} \eta^{k}-e^{(2\eta^k-2\eta^{k-1})}+1 \right)
\end{equation}
with $\mu \in (t,x)$. Then it is straightforward to verify that:
\[
 \de_{\mu}\frac{\de\mathcal{L}}{\de (\de_\mu\eta^k)
}-\frac{\de\mathcal{L}}{\de\eta^k}=0 \quad \Longrightarrow~~
 \de^2_t \eta^k- \de_x^2 \eta^k=2 e^{(2\eta^{k+1}-2\eta^{k})}-2 e^{(2\eta^k-2\eta^{k-1})}.
\]
Typically the further condition $\sum_k \eta^k=0$ is imposed
on the fields\footnote{This condition has the effect of excluding
a zero mode in the \textquotedblleft mass spectrum\textquotedblright 
of the theory:
\[
 m^2_{n}=4\sin^2(\pi n/N),\quad n=0,1,\dots,N-1\,.
\]}, which results in the affine Toda field theory based on the
$a^{(1)}_{N-1}$ affine Dynkin diagram.
For $N=2$, 
with $\eta^1=-\eta^2=\eta$,
the equations then reduce to
\begin{equation}
  \eta_{tt}-\eta_{xx}+4\sinh(2\eta)=0 \quad  \text{\textbf{(the 
sinh-Gordon equation)}}.
\end{equation}
For $N=3$, and with $\eta^3=-\eta^1-\eta^2$,
\begin{equation}\label{BD-grezzo}
 \begin{aligned}
  \eta^1_{tt}-\eta^1_{xx}&+2e^{(2\eta^1-2\eta^2)}-2 e^{(-4\eta^1-2\eta^2)}=0 ,\\
  \eta^2_{tt}-\eta^2_{xx}&+2e^{(4\eta^1+2\eta^2)}-2\,e^{(2\,\eta^1-2\,\eta^2\,)}\,=0,
 \end{aligned}
\end{equation}
which are the equations of motion for the $a_2^{(1)}$ affine Toda
field theory. The $\ZZ_2$ symmetry of the $a_2^{(1)}$ affine Dynkin
diagram allows a further symmetry to be imposed,
namely $\eta^1=-\eta^2=\eta$. 
Then \eqref{BD-grezzo} becomes
\begin{equation}\label{BD}
\eta_{tt}-\eta_{xx}+ 2e^{4\eta}- 2e^{-2\eta}=0 
\quad \text{\textbf{(the Bullough-Dodd equation)}}.
\end{equation}

Equation \eqref{BD} can also be recovered directly
from the compatibility of $3
\times 3$ matrix operators which are defined as \cite{Toda}
\begin{align}\label{BD-sys}
 T&=\de_x +
\begin{pmatrix}
 \eta_t                   & - i \la^{-1}e^{-\eta}  &  i \la e^{2\eta} \\
  i \la e^{-\eta}      & 0                         & - i \la^{-1}e^{-\eta}                     \\
 - i \la^{-1}e^{2\eta} & i \la e^{-\eta}        &-\eta_t
\end{pmatrix} \\
X&=\de_t +
\begin{pmatrix}
 \eta_x                   &  i \la^{-1}e^{-\eta}    &  i \la e^{2\eta} \\
  i \la e^{-\eta}      & 0                          &  i \la^{-1}e^{-\eta}                     \\
  i \la^{-1}e^{2\eta}  &  i \la e^{-\eta}        &-\eta_x
\end{pmatrix}.
\end{align}
If we were to introduce light-cone coordinates as
\begin{equation}
 z=x+t \qquad \bz=x-t,
\end{equation}
and redefine $\eta\rightarrow \eta/2$,  \eqref{BD} would become
\begin{equation}\label{BD-lc}
 \de_z\de_{\bz}\eta(z,\bz) + e^{-\eta(z,\bz)}-e^{2\eta(z,\bz)}=0\,.
\end{equation}
The nonlinear equation \eqref{BD-lc} is the starting point for the
construction of objects which are in correspondence with the quantum
world. However it is important to stress that, for this purpose, the
coordinates $z$ and $\bz$ should be considered as living in an
`auxillary space' which is {\em not}\/ the same as the space on
which the quantum field theory under discussion will be defined. 
In fact it
will be better to adopt a Euclidean metric on this auxillary space so
that $z$ and $\bz$ are complex conjugates of each other, though
this will often be relaxed and $z$ and $\bz$ 
regarded as independent complex variables.
To pave the way to the correspondence, it 
is also convenient to adopt a modified version
of \eqref{BD-lc}. This will be referred to as the modified
Bullough-Dodd (mBD) equation and is
\begin{equation}\label{mBD}
 \de_z\de_{\bz}\eta+e^{-\eta(z,\bz)}-p(z,s^{3\al})p(\bz,s^{3\al})
e^{2\eta(z,\bz)}=0,
\quad \text{\textbf{(the mBD equation)}}
\end{equation}
where $\al>0$ and the function $p(z,s^{3\al})$ is defined to be
\begin{equation}
 p(z,s^{3\al})=z^{3\al}-s^{3\al}, 
\end{equation}
and $s$ will turn out to take the role of
a scale parameter.
Equations \eqref{BD-lc} and \eqref{mBD} are related by a change of
variables and a redefinition of the field, since, defining
\begin{equation}
 w(z)=\int^{z} \sqrt{p(z')}dz' \quad \text{and} \quad
\bar{w}(\bz)=\int^{\bz} \sqrt{p(\bz')}d\bz',
\end{equation}
and shifting the field as $\eta \rightarrow \eta +
\frac1{3}\ln(p(z)p(\bz))$, the mBD equation reduces to the
Bullough-Dodd equation.
The mBD model has in general no explicit rotational symmetry. Instead
it has a discrete symmetry dictated by the form of
the `potential' function $p(z)$:
\begin{equation}
 z\rightarrow e^{\frac{2\pi i }{3\al}}z,\quad \bz\rightarrow e^{-\frac{2\pi i }{3\al}}\bz.
\end{equation}

The solutions of the mBD equation \eqref{mBD} which will be relevant for
the current  analysis must respect this symmetry, be continuous at
every finite nonzero $z$, $\bz$,  and grow more slowly than exponentially 
at $z,\bz\rightarrow\infty$. As in \cite{Lukyanov:2010rn} for the
(modified) sinh-Gordon case, we will consider a
one-parameter family of such solutions, characterized
by a logarithmic behaviour near the origin.  Before listing in detail
the exact conditions which are required, polar coordinates can be
introduced to show these properties more neatly:
\begin{equation}\label{polarc}
 z=\rho e^{ i \phi},\qquad \bz=\rho e^{- i \phi}.
\end{equation}
In most the following $z$ and $\bz$ will be treated as
independent complex variables,  so the solution $\eta$
will be a function of the independent variables
$(\rho,\phi)$, according to \eqref{polarc}.

It is now possible to list the properties of the sought-after
solutions:
\begin{enumerate}[(i)]
 \item periodicity:
\begin{equation}\label{eta_period}
 \eta(\rho,\phi+\frac{2\pi}{3\al})=\eta(\rho,\phi);
\end{equation}
or, better, the solutions $\eta(\rho,\phi)$ are single-valued
functions on a cone with the apex angle $\frac{2\pi}{3\al}$,
\begin{equation}\label{cone}
 \mathcal{C}_{\frac{2\pi}{3\al}}\;:\;\quad \phi\sim\phi+\frac{2\pi}{3\al},\quad 0\le\rho<\infty;
\end{equation}
\item the solutions
$\eta(\rho,\phi)$ are real-valued for real $\rho$ and $\phi$,
and finite everywhere on the cone $\mathcal{C}_{\frac{2\pi}{3\al}}$, except
for the apex $\rho=0$;
\item large-$\rho$ asymptotic:
\begin{equation} \label{L-as}
 \eta(\rho,\phi)=-2\al\ln\rho + o(1)\quad \text{as} \quad
\rho\rightarrow \infty;
\end{equation}
\item small-$\rho$ asymptotic:
\begin{equation}\label{S-as}
 \eta(\rho,\phi)=-2 g \ln(\rho) + O(1) \quad \text{for} \quad -1<g<1/2.
\end{equation}
\end{enumerate}
\nobreak
The factors $-2\al$ and $-2g$ in \eqref{L-as} and \eqref{S-as} have 
been chosen, for future convenience, to be consistent with 
\cite{Dorey:1999pv} and \cite{Lukyanov:2010rn}.
\goodbreak

As in \cite{Lukyanov:2010rn}, one can
start from equation \eqref{S-as} and develop an
expansion for $z,\bz \sim 0$ of the form
\begin{align}\label{eta-exp}
 \eta\,= \, & -g\ln(z\bz)+\eta_0+
\sum_{k=1}^{\infty}\gamma_k\bigl(z^{3Mk}+\bz^{3Mk}\bigr) \nonumber\\
&~~~ -\frac{e^{-\eta_0}}{(g+1)^2}(z\bz)^{g+1}-
\frac{s^{6\al}e^{2\eta_0}}{(-2g+1)^2}(z\bz)^{-2g+1}+\dots
\end{align}
where $\eta_0$ $\gamma_k$ are integration constants, all remaining 
terms omitted in the expansion being uniquely
determined once these constants are given. These constants are not
new parameters, but should be fixed by demanding  the
consistency of this expansion with the remaining conditions
(i)-(iii) above.

The expansion \eqref{eta-exp} remains valid with $z$ and $\bz$ 
regarded as independent complex variables  and, for later analysis,
it will be useful to record the form of (\ref{S-as}) in the so-called
\textit{light-cone} limit $\bz\rightarrow0$ (with fixed $z$):
\begin{equation}\label{lcl-eta}
 \eta\sim -g \ln(z\bz)+\eta_0+\gamma(z)
\end{equation}
where $\gamma(z)=\sum_{k=1}^{\infty}\gamma_kz^{3Mk}$.

At this stage it is possible to define the linear problem associated
to the mBD equation
\begin{equation}
\label{mBD-sys-simb}
 \bm{D}\bm{\Psi}=0,\qquad \bm{\bar{D}}\bm{\Psi}=0,
\end{equation}
where $\bm{D}$ and $\bm{\bar{D}}$ are components of an \textit{sl(3)}
connection and can be found by rearranging, in light-cone
coordinates, the matrix operators \eqref{BD-sys} with an appropriate
redefinition of the  spectral parameter $\la$ and after the
introduction of  the potential $p(z)$:
\begin{align} 
 \bm{D}=&\de_z+
\begin{pmatrix}\label{mBD-sys1}
 \frac1{2}\de_z\eta        & 0                       & \la e^{\eta} p(z)   \\
 -\la e^{-\frac1{2}\eta}   & 0                       & 0                 \\
 0                         & -\la e^{-\frac1{2}\eta} & -\frac1{2}\de_z \eta
\end{pmatrix}, \\[2pt]
\bm{\bar{D}}=&\de_{\bz}+
\begin{pmatrix}\label{mBD-sys2}
 -\frac1{2}\de_{\bz}\eta   & \la^{-1} e^{-\frac1{2}\eta} & 0                           \\
 0                         & 0                       & \la^{-1}e^{-\frac1{2}\eta} \\
 \la^{-1}p(\bz)e^\eta     & 0                       & \frac1{2}\de_{\bz}\eta
\end{pmatrix}.
\end{align}
It is possible to deal directly with this system of six equations on $\bm{\Psi}$  to get information
about the solutions and their asymptotics, but it turns out to be
simpler to deal with a reduction of \eqref{mBD-sys1},
\eqref{mBD-sys2}
to two third-order differential equations. Moreover this step will
be necessary for the continuation of the analysis and to show
the connection between the original nonlinear problem and a
previously-studied
spectral problem.  This reduction can be implemented defining a vector
solution to equations \eqref{mBD-sys1}, \eqref{mBD-sys2}%
\footnote{This is simply recovered solving
each of the systems \eqref{mBD-sys-simb} respect to two out of three
components of the vector $\bm{\Psi}$.}:
\begin{equation}\label{mBD-vecsol}
 \bm{\Psi}=
\begin{pmatrix}
\la^{-\frac1{2}}e^{\eta/2}\de_z(e^{\eta}\de_z(e^{-\eta}\psi)) \\
-\la^{\frac1{2}}e^{\eta}\de_z(e^{-\eta}\psi)                  \\
\la^{\frac3{2}}e^{-\eta/2}\psi
\end{pmatrix}=
\begin{pmatrix}
 \la^{-\frac3{2}}e^{-\eta/2}\bar{\psi}                            \\
-\la^{-\frac1{2}}e^{\eta}\de_{\bz}(e^{-\eta}\bar{\psi})           \\
\la^{\frac1{2}}e^{\eta/2}\de_{\bz}(e^{\eta}\de_{\bz}(e^{-\eta}\bar{\psi}))
\end{pmatrix}.
\end{equation}
 The equality between each row of the two parentheses in
\eqref{mBD-vecsol} must always hold, and will be useful to find
large-$\rho$ asymptotics of \eqref{mBD-sys-simb}. Applying
$\mathbf{D}$ and $\mathbf{\bar{D}}$ to the first and the second
vectors
in \eqref{mBD-vecsol} respectively, it is simple to find the system of
two third-order linear differential equations which constitutes a
compatibility condition on $\psi$ and $\bar{\psi}$ so that $\bm{\Psi}$
really is a solution of \eqref{mBD-sys-simb}. The two equations are:
\begin{align}
\de^3_{z}\psi&-((\de_z\eta)^2+2\de^2_{z}\eta)\de_z\psi+(\la^3\;\;p(z)-\de_z\eta\de^2_{z}\eta
-\de^3_{z}\eta)\psi=0 \label{3ord1};\\
\de^3_{\bz}\bar{\psi}&-((\de_{\bz}\eta)^2+2\de^2_{\bz}\eta)\de_{\bz}\bar{\psi}+
(\la^{-3}p(\bz)-\de_{\bz}\eta\de^2_{\bz}\eta-\de^3_{\bz}\eta)\bar{\psi}=0. \label{3ord2}
\end{align}

The next step is to determine the $\rho\rightarrow 0$ asymptotics of
solutions to \eqref{mBD-sys-simb}. To do this it is convenient
to focus attention on eq.~\eqref{3ord1}, though the result
would not change if eq.~\eqref{3ord2} was taken as the starting
point instead. This equation is entirely in $z$, except for the
appearance of $\bz$ in
$\eta(z,\bz)$ that, for now, can be considered as a simple
parametric dependence. Finding an asymptotic solution $\psi(z)$ 
enables a solution $\bm{\varPsi}$ of
\eqref{mBD-sys-simb} for $\rho\rightarrow 0$ to be determined through 
\eqref{mBD-vecsol}.  Substituting the asymptotic form 
\eqref{S-as} of $\eta$
into \eqref{3ord1} and considering the $\rho\rightarrow0$
limit (actually $z\rightarrow0$) brings equation \eqref{3ord1} into the
form
\[
 \de^3_{z}\psi-\frac1{z^2}g(g+2)\de_z\psi+\frac1{z^3}g(g+2)\psi=0,
\]
and seeking solutions of the form $\psi=z^{\mu}+\dots$ the problem is
reduced to the solution of the following  indicial equation:
\[
 (\mu-1)[\mu(\mu-2)-g(g+2)]=0.
\]
This equation gives three different leading behaviours in
the $z\rightarrow0$ limit:
\begin{equation}\label{psi0}
\quad \psi_+\sim z^{-g}, \quad\psi_0\sim z, \quad \psi_-\sim z^{g+2}.
\end{equation}
These expressions give the leading behaviours and can be adjusted 
with appropriate multiplicative  constants;
once inserted
into \eqref{mBD-vecsol} in the usual $\rho\rightarrow0$ limit, they
provide the three asymptotic solutions $\Psi$ to
\eqref{mBD-sys-simb}. They are:
\begin{equation} \label{Psi0}
 \bm{\varPsi}_+\sim
\begin{pmatrix}
0               \\
0               \\
e^{( -i \phi-\te)g}
\end{pmatrix}, \quad
\bm{\varPsi}_0\sim
\begin{pmatrix}
0 \\
1 \\
0
\end{pmatrix}, \quad
\bm{\varPsi}_-\sim
\begin{pmatrix}
e^{( i \phi+\te)g} \\
0                        \\
0
\end{pmatrix} \qquad (\rho\rightarrow 0)
\end{equation}
with $\la=e^{\te}$. 
In
order to fix the values of the constants introduced above a particular
symmetry condition has been considered. In fact, although
$\eta(\rho,\phi)$ is a single-valued function on the cone
\eqref{cone}, the connection components \eqref{mBD-sys1} and
\eqref{mBD-sys2} are not.  Instead, the linear problem
\eqref{mBD-sys-simb} is invariant with respect to the transformation
\begin{equation}
 \Omega:\qquad \phi\rightarrow \phi+\frac{2\pi}{3\al},\quad \te\rightarrow \te-\frac{2\pi i }{3\al},
\end{equation}
involving the shift of the spectral parameter $\te$ (from now on $\te$
and $\la$ will be referred as spectral parameters
interchangeably,  keeping in mind that $\la=e^{\te}$).
Imposing this symmetry, it is straightforward to find \eqref{Psi0}.

On the other hand it is also possible to give the large-$\rho$
asymptotics, and for this it is best to use
a semiclassical approximation. As will be explained more formally
below, assuming that $\te$ is real,
a generic solution of \eqref{3ord1} grows exponentially at
$\rho\rightarrow\infty$, but 
\cite{Sha,Dorey:1999pv}
there are special (`subdominant')
solutions that decay
in the sector 
\begin{equation}
 -\frac{4\pi}{3\al+3}<\phi<\frac{4\pi}{3\al+3}\,.
\end{equation}
In order to build such a solution to \eqref{mBD-sys-simb},
the first step is  the specification of a $\rho\rightarrow\infty$
asymptotic for \eqref{3ord1}, just as done before for the solution in
the region near $ \rho=0$, again remembering to substitute the
large-$\rho$ solution of \eqref{mBD}.  In this particular limit
equation \eqref{3ord1} reduces  to a sum of a third order derivative
and a \textquoteleft potential\textquoteright \;$Q(z)$ which will fix
the leading term of the subdominant WKB solution:
\begin{equation}\label{wkb-pot}
 Q(z)=e^{3\te}(z^{3\al}-s^{3\al}).
\end{equation}
It follows from \eqref{wkb-pot} that the WKB-like solution has, for $\al > 1/2$,  the form:
\begin{equation}\label{wkb-sol.grezza1}
 \psi\sim c_1z^{-\al}\exp \left(-\frac{z^{\al+1}}{\al+1}e^{\te}+f(\bz) \right),
\end{equation}
where $c_1$ is an arbitrary constant and the term $f(\bz)$ in the
exponential carries the dependence on $\bz$ of $\psi$.
To retrieve a more exact form of \eqref{wkb-sol.grezza1}, it is
convenient to seek a solution of \eqref{3ord2} of the
form
\begin{equation}\label{wkb-sol.grezza2}
 \bar{\psi}\sim c_2\bz^{-\al}\exp \left(-\frac{\bz^{\al+1}}{\al+1}e^{-\te}+g(z) \right)
\end{equation}
and then by inserting both the results in \eqref{mBD-vecsol} the
compatibility between the two solutions will fix
these integration-generated functions. This yields:
\begin{equation}
 c_2=c_1e^{\te}=ce^{\te},\quad f(\bz)=-\frac{\bz^{\al+1}}{\al+1}e^{-\te},\quad g(z)=-\frac{z^{\al+1}}{\al+1}e^{\te},
\end{equation}
and the final semiclassical  solution of \eqref{3ord1} (in polar coordinates) is
\begin{equation}\label{wkb-sol-3ord}
 \psi\stackrel{\text{\tiny{WKB}}}{\sim}ce^{-\te}z^{-\al}\exp \left(-\frac{2\rho^{\al+1}}{\al+1}\cosh(\te+ i (\al+1)\phi) \right).
\end{equation}
By inserting \eqref{wkb-sol-3ord} into the first of
\eqref{mBD-vecsol} it is then straightforward to obtain
\begin{equation}
 \Psi\stackrel{\text{\tiny{WKB}}}{\sim} e^\frac{\te}{2}
\begin{pmatrix}
e^{ i \phi\al}  \\
1                    \\
e^{- i \phi\al}
\end{pmatrix}
\exp\left (-\frac{2\rho^{\al+1}}{\al+1}\cosh(\te+ i (\al+1)\phi)\right),\quad \rho\rightarrow\infty.
\end{equation}
Since the functions
$\{\varPsi_{+},\varPsi_{0},\varPsi_{-}\}$ form a basis in the space of
solutions of linear problem \eqref{mBD-sys-simb},
a linear relation of the following form must hold:
\begin{equation}
 \Psi=Q^+(\te,s)\varPsi_++Q^0(\te,s)\varPsi_0+Q^-(\te,s)\varPsi_-,
\end{equation}
where the coefficients $Q^{\pm,0}$, which are independent of the variables 
$\rho$ and $\phi$,   are functions of the spectral parameter
$\te$ and of $s$,  as well as of the parameter $g$ (this parameter has
been temporarily omitted). These coefficients will coincide
(up to an overall constant) with $Q$-functions of a 2d massive quantum
field theory related to the Lie algebra $A_2$, or $su(3)$.

\section{The conformal limit}
\label{conflim-subsec}
Before proceeding with the advertised correspondence, it is
useful to work a little more on equation \eqref{3ord1}.
The problem studied up to now has an important connection
with a known spectral problem of a particular third-order differential
equation. In order to build an appropriate analysis and to understand
where the various steps come from,
\eqref{3ord1} will be reduced to this spectral problem and a brief 
review of previously-obtained results will be given.

Recall that equation \eqref{3ord1} is of the form
\[
 \de^3_{z}\psi-\left((\de_z\eta)^2+2\de^2_{z}\eta\right)\de_z\psi+
\left(\la^3\;p(z, s^{3\al} )-\de_z\eta\de^2_{z}\eta
-\de^3_{z}\eta\right)\psi=0,
\]
and, as was said above, 
$\bz$ plays the role of a simple parameter, so it is possible to take
the so-called
light-cone limit $\bz\rightarrow0$ (or, if \eqref{3ord2} had
been chosen as starting point then the light-cone
limit would have involved $z$ instead of $\bz$) in which $\eta$
assumes the form \eqref{lcl-eta}. After that, the limit
$z\sim s\rightarrow0$, $\te\rightarrow+\infty$ can be 
taken, with the combinations
\begin{equation}\label{conf-limit}
 x=e^{\frac{\te}{\al+1}}z,\quad E=s^{3\al}e^{\frac{3\te\al}{\al+1}}
\end{equation}
kept finite while
\begin{equation}\label{conf-limit1}
 \tilde{x}=e^{-\frac{\te}{\al+1}} \bar{z} \rightarrow 0,\quad \tilde{E}=s^{3\al}e^{-\frac{3\te\al}{\al+1}} \rightarrow 0.
\end{equation}
This limit is a particular scaling limit because $z$, $s$
are sent to zero and $\te$ to infinity, but they are rearranged in
combinations such that the new \textquoteleft 
variables\textquoteright
\; $x$ and $E$  do not diverge or collapse.  This
is an
important turning point of the theory, since it transforms the
correspondence between an ODE and a massive QFT into a correspondence
with a conformal field theory 
\cite{Dorey:1999pv} (which is massless).  For this reason
the limit \eqref{conf-limit1} can be called the \textit{conformal
limit}, with $s$ (or, more precisely, $s^{M+1}$) taking the role of
the mass scale.

Inserting \eqref{conf-limit} in \eqref{3ord1} and taking the limits
just discussed,
equation \eqref{3ord1} reduces to
\begin{equation}\label{cong-eq}
 \left[\de^3_{x}-g(g+2)\left(\frac1{x^2}\de_{x}-\frac1{x^3}\right)+p(x,E)\right]y(x,E,g)=0,
\end{equation}
where $y$ has been used instead of $\psi$ to avoid confusion among the
various solutions.
Motivated by the results of \cite{Dorey:1999pv} and \cite{Sha},  
a generalization of a theorem due to Sibuya for second-order equations
can now be proposed:

\begin{conjecture}
\label{sibuya-theo}
Equation \eqref{cong-eq} has a solution $y(x,E,g)$ with the 
following properties:
\begin{enumerate}[(i)]
 \item $y$ is an entire function of (x,E) though, due to the branch point in the potential at $x=0$, $x$ must in general
be considered to live on a suitable cover of the punctured complex plane;
\item $y$, $y'=\de_{x} y$ and $y''=\de^2_{x}y $ admit, for $\al>1/2$, the asymptotic representations
\begin{equation} \label{sibuya-form}
 y  \sim x^{-\al}e^{-\frac1{\al+1}x^{\al+1}},\quad
 y' \sim -e^{-\frac1{\al+1}x^{\al+1}},\quad
 y''\sim x^{-\al}e^{-\frac1{\al+1}x^{\al+1}},
\end{equation}
as $x\rightarrow\infty$ in the sector
\begin{equation}
 |\arg(x)|<\frac{4\pi}{3\al+3}\,;
\end{equation}
\item $y$ is uniquely fixed by the properties i) and ii).
\end{enumerate}
\end{conjecture}

The behaviour of the solution \eqref{sibuya-form}  follows from
the more general formula
\begin{equation}
 y(x,E)\sim Q(x,E)^{-1/3}\exp\left(-\int_{x_0}^{x}Q(x,t)^{1/3}dt\right),
\end{equation}
where $y$ solves a generic equation $\de^3_{x} y+Q(x,E)y=0$. This is the 
analogue of a WKB approximation for a solution of a
Schr\"{o}dinger equation.
It is possible to define rotated solutions, by analogy with
\cite{Dorey:2007zx}, in order to construct bases of solutions for
\eqref{conf-limit}. For general values of $k$, define
\begin{equation}\label{rot-sol-conf}
 y_k(x,E,g)=\omega^ky(\omega^{-k}x,\omega^{-3\al k}E,g),
\end{equation}
with
\begin{equation}
 \omega=\exp\left(\frac{2\pi i }{3\al+3}\right).
\end{equation}
Substituting in, $y_k$ solves
\begin{equation}\label{conf-limit-rot}
\de^3_{x}  y _k-g(g+2)\left(\frac1{x^2}\de_{x} -\frac{1}{x^3}\right)y_k+e^{-2k\pi i }
p(x,E)y_k=0,
\end{equation}
and so for $k\in\mathbb{Z}$, \eqref{rot-sol-conf} provides a potentially-new 
solution to \eqref{conf-limit}. However
 for now it is convenient to leave $k$ arbitrary since fractional
values will also be needed. We also
define the \textit{Stokes sectors}
\begin{equation}\label{stokes}
 \mathcal{S}_k\; : \quad \left|\arg(x)-\frac{2\pi k}{3\al+3}\right|<\frac{\pi}{3\al+3}.
\end{equation}
Some discussion of the dominance and subdominance of solutions
is also required, since, for a third-order problem, this is a little
more complicated than the second-order case. For this we largely
follow \cite{Dorey:2007zx}. First, the behaviour specified by
conjecture
\ref{sibuya-theo} lies in the sectors
$\mathcal{S}_{-\frac3{2}}\cup\mathcal{S}_{-\frac1{2}}
\cup\mathcal{S}_{\frac1{2}}\cup
\mathcal{S}_{\frac3{2}}$. Furthermore, in general, there are three
types of asymptotic solution for large $|x|$. Aside the
well-behaved leading term $x^{-\al}\exp(-x^{\al+1}/(\al+1))$, there
are also solutions that behave as
$x^{-\al}$ $\exp(e^{\pm\pi i /3}x^{\al+1}/(\al+1))$. (This can be
traced to the three third roots of $-1$, which are
$-1$, $e^{\pi i /3}$ and $e^{-\pi i /3})$. According to the sector is
being considered, either one or two of these solutions tend to zero for large
$|x|$. The \textit{subdominant} solution in any given sector
will be defined to be the one
that tends to zero fastest in that sector (it might also be called
maximally subdominant). Finally, $y_k$, up to
overall scalar factor, is characterized as the solution of
\eqref{conf-limit} which is subdominant in the sector $\mathcal{S}_k$.  
The asymptotic \eqref{sibuya-form} along with the definition
\eqref{rot-sol-conf} imply
\begin{equation}
\begin{aligned}\label{rot-sol-asy}
 y_k\sim
\omega^{k(\al+1)}x^{-\al}e^{-\frac{x^{\al+1}}{\al+1}\omega^{-(\al+1)k}},& 
\quad
y'_k\sim-e^{-\frac{x^{\al+1}}{\al+1}\omega^{-(\al+1)k}}, \\
y''_k\sim\omega^{-k(\al+1)}x^{\al}e^{-\frac{x^{\al+1}}{\al+1}\omega^{-(\al+1)k}},&
\end{aligned}
\end{equation}
for $|x|\rightarrow\infty$ with
\begin{equation}\label{x-region}
 x\in\mathcal{S}_{k-\frac3{2}}\cup\mathcal{S}_{k-\frac1{2}}
\cup\mathcal{S}_{k+\frac1{2}}\cup\mathcal{S}_{k+\frac3{2}}.
\end{equation}
In the region formed by the two sectors
$\mathcal{S}_{k+\frac1{2}}\cup\mathcal{S}_{k+\frac3{2}}$, the three
asymptotics $y_k$, $y_{k+1}$ and $y_{k+2}$ live together and comparing
their forms, given by \eqref{rot-sol-asy}, they turn out to be
independent of each other.  A further proof of this fact can be built
defining the \textit{Wronskian} of the three solutions:
\begin{equation}
 W_{k_1,k_2,k_3}=W[y_{k_1},y_{k_2},y_{k_3}]
\end{equation}
where the $3 \times 3$ Wronskian is defined as
\begin{equation}
 W[f,g,h]=det
\begin{pmatrix}
 f   & f'   & f''    \\
 g   & g'   & g''    \\
 h   & h'   & h''
\end{pmatrix}.
\end{equation}
For $f$, $g$ and $h$ solving \eqref{conf-limit}, $W[f,g,h]$ is independent
of $x$, and $f$, $g$ and $h$ are linearly
independent if and only if $W[f,g,h]\neq0$.
In order to demonstrate the independence of $y_k$, $y_{k+1}$ and
$y_{k+2}$, it is helpful first to compute
$(k_1,k_2,k_3)=(-1,0,1)$
\begin{equation}
 W_{-1,0,1}=-3 i \sqrt{3}
\end{equation}
and then, using the general result
\begin{equation}
 W_{k_1+b,k_2+b,k_3+b}(E)=W_{k_1,k_2,k_3}(\omega^{-3\al b}E),
\end{equation}
we see that $W_{k,k+1,k+2}$ is always nonzero, thus confirming
the independence of $\{y_k,y_{k+1},y_{k+2}\}$.

The final step for this section is to show how to build new solutions
to \eqref{conf-limit} using the $y_k$'s.
Supposing that $k_1$ and $k_2$ differ by an integer so that
$e^{-2k_1\pi i }=e^{-2k_2\pi i }=e^{-2k\pi i }$), it can be 
checked that the function
\begin{equation}\label{zconf-sol}
 z_{k_1,k_2}(x,E,g)=y_{k_1}y'_{k_2}-y_{k_2}y'_{k_1},
\end{equation}
which is actually a $2 \times 2$ Wronskian, provides a solution of
\begin{equation}\label{zconf-limit-rot}
\de^3_{x} z_{k_1,k_2}-g(g+2)\left(\frac1{x^2} \de_{x}-\frac{1}{x^3}\right)z_{k_1,k_2}-e^{-2k\pi i }
(x^{3\al}-E)z_{k_1,k_2}\,.
\end{equation}
Equation \eqref{zconf-limit-rot} is the adjoint of \eqref{conf-limit-rot} and if $k$ is shifted by a half-integer, then
$z_{k_1,k_2}$ becomes a solution of the original equation \eqref{conf-limit-rot}
\begin{equation}\label{zconf-limit-rot2}
 \de^3_{x}  z_{k_1+\frac1{2},k_2+\frac{1}{2}}-g(g+2)\left(\frac1{x^2} \de_{x} -\frac1{x^3}\right)
z_{k_1+\frac1{2},k_2+\frac1{2}}+e^{-2k\pi i }(x^{3\al}-E)z_{k_1+\frac1{2},k_2+\frac1{2}}.
\end{equation}
For $|k_1-k_2|<3$, the regions \eqref{x-region} for $k_1=k_2=k$ overlap, and it is possible to get an asymptotic for
$z_{k_1,k_2}$ from \eqref{rot-sol-asy}. In particular, for $k=1,2,3$
we have
\begin{equation}\label{zrot-sol-asy}
 z_{-k/2,k/2}(x,E,g)\sim2 i \sin(\pi k/3)x^{-\al}e^{-2\cos(\pi k/3)\frac1{\al+1}x^{\al+1}},\quad
x\rightarrow\infty.
\end{equation}
Considering \eqref{zconf-limit-rot} and \eqref{zrot-sol-asy}, for $k=1$ it follows, by uniqueness of solutions, that
\begin{equation}\label{conf-func-pin}
 z_{-\frac1{2},\frac1{2}}(x,E,g)= i \sqrt{3}y(x,E,g).
\end{equation}
This result is limited to the $(k=1)$-case, as
for other values of the parameter is not possible to retrieve enough
information to uniquely pin the function down. On the other hand
\eqref{conf-func-pin} is useful to determine the
\textit{Bethe Ansatz equations}, 
but this will be shown in the next section, starting from the original problem
\eqref{3ord1}.
Now that the analyticity properties of solutions of \eqref{conf-limit}
have been studied, it is possible to return
to \eqref{3ord1} and apply to that, with the appropriate caution, all
the results that have been found in this section.

\section{Bethe Ansatz equations}
\label{BAE-sec}
In this section, Bethe Ansatz equations will be
built starting from certain functional relations involving
the coefficients $Q^{\pm}$ already introduced at the end of \secref{BDdef-sec}. 
In order to get there it is necessary to apply the same analysis
described in \secref{conflim-subsec} to reproduce similar
results for the original equation \eqref{3ord1}, which is repeated
below for convenience:
\begin{equation}\label{E-eq}
 \de^3_{x}\psi-\bigl((\de_x\eta)^2+2\de^2_{x}\eta
\bigr)\de_x\psi+\bigl(p(x,E)-\de_x\eta\de^2_{x}\eta
-\de^3_{x}\eta \bigr)\psi =0,
\end{equation}
\begin{equation}\label{E-bar-eq}
 \de^3_{\tilde x}\psi-\bigl((\de_x\eta)^2+2\de^2_{\tilde x}\eta
\bigr)\de_x \bar{\psi}+\bigl(p(\tilde x,\tilde E)-\de_x\eta\de^2_{\tilde x}\eta
-\de^3_{\tilde x}\eta \bigr)\bar{\psi} =0,
\end{equation}
where
\begin{equation}
x= z e^{\theta/(\al+1)},~~ \tilde{x}= \bar{z} e^{-\theta/(\al+1)},~~
E= s^{3 \al} e^{3 \theta \al/(\al+1)},~~ \tilde{E}= s^{3 \al} e^{-3\theta \al /(\al+1)},~~
\end{equation}
and
\begin{equation}
\psi \equiv \psi(x,\tilde{x}, E, \tilde{E}, g),~~
\bar{\psi} \equiv \bar{\psi}(x,\tilde{x}, E, \tilde{E}, g).
\end{equation}

First of all, the rotated solutions, analogues to
\eqref{rot-sol-conf}, have to be defined. In order to do this, the
periodicity of $\eta(\rho,\phi)$ has to be exploited since it is the
only information, besides the asymptotic behaviours, which is known.
It turns out that the right form is
\begin{equation}\label{rot-sol}
\psi_k(x,\tilde{x}, E, \tilde{E}, g)=\om^{k}\psi(\om^{-k} x, \om^{k}
\tilde{x},  \om^{-3 \al k} E,
\om^{3 \al k} \tilde{E},g ),
\end{equation}
where
\begin{equation}
 \om=\exp\left(\frac{2\pi i }{3\al+3}\right).
\end{equation}

The function \eqref{rot-sol} solves
\begin{equation}\label{rot-eq}
\de^3_{x}\psi_k-
\bigl((\de_x\eta)^2+2\de^2_{x}\eta\bigr) \de_x \psi_k+
\bigl(e^{-2k\pi i } p(x,E)-\de_x\eta\de^2_{x}\eta-\de^3_{x}\eta\bigr)\psi_k
=0,
\end{equation}
where, analogously to \eqref{conf-limit-rot}, a $e^{-2k\pi i }$-term
appears. In general arbitrary integer or
half-integer values for $k$ will be considered, 
and the definition of Stokes sectors
remains unchanged (see \eqref{stokes}).  Following the discussion of
\secref{conflim-subsec}, it is straightforward to find the $|x|
\rightarrow \infty$ behaviour of the rotated solutions
\eqref{rot-sol}, (keeping in mind that $\tilde{x}$ goes to $\om^k
\tilde{x} $) and setting
\begin{equation}
c=e^{\theta/(\al+1)}
\end{equation}
\begin{equation}\label{asy-sol}
\begin{aligned}
 \psi_k\sim\om^{k(\al+1)} x^{-\al}\exp\left[
-\frac{x^{\al+1}}{\al+1} \om^{-k(\al+1)}-\frac{\tilde{x}^{\al+1}}{\al+1} \om^{k(\al+1)}\right],& \\
\psi'_k\sim-\exp\left[
-\frac{x^{\al+1}}{\al+1} \om^{-k(\al+1)}-\frac{\tilde{x}^{\al+1}}{\al+1}\om^{k(\al+1)}\right],& \\
\psi''_k\sim\om^{-k(\al+1)} x^{\al}\exp\left[
-\frac{x^{\al+1}}{\al+1} \om^{-k(\al+1)}-\frac{\tilde{x}^{\al+1}}{\al+1}\om^{k(\al+1)}\right].&
\end{aligned}
\end{equation}
(the derivatives are on $x$ only.)

Now, the three generic solutions $\{\psi_k,\psi_{k+1},\psi_{k+2}\}$
form a basis for the solutions of eq.\eqref{3ord1}. Again, the proof
of this property comes from the evaluation of the Wronskian for
$(k_1,k_2,k_3)=(-1,0,1)$
\begin{equation}
 W_{-1,0,1}=W[\psi_{-1},\psi_{0},\psi_{1}]=-3 i \sqrt{3},
\end{equation}
which shows that these three solutions are independent.

The next step is to define the analogues of the functions \eqref{zconf-sol}
that, here, will be denoted as $u_{k_1,k_2}$ so as to avoid confusion 
with the variable $z$. They read as
\begin{equation}\label{u-sol-rot}
 u_{k_1,k_2}(x,\tilde{x}, E, \tilde{E},g)=[\psi_{k_1}\psi'_{k_2}-\psi_{k_2}\psi'_{k_1}](x,\tilde{x}, E, \tilde{E},g),
\end{equation}
where, again, the $\bz$-dependence has been omitted. The solutions \eqref{u-sol-rot} solve
\begin{equation}
\de^3_x u_{k_1,k_2}-\left((\de_x\eta)^2+2\de^2_{x}\eta\right) \de_x u_{k_1,k_2}-\left(e^{-2\pi i  k}p(x,E)
+\de_x\eta\de^2_{x}\eta+\de^3_{x}\eta \right)u_{k_1,k_2}=0,
\end{equation}
and, also in this case, it is possible to evaluate \eqref{u-sol-rot} for $k_1=-k_2=-\frac{k}{2}$

\begin{equation}
 u_{-k/2,k/2} \sim 2  i  \sin(\frac{\pi}3k)  x^{-\al} \exp\left[-2\cos(\frac{\pi}3k) \left(
\frac{x^{\al+1}}{\al+1} +\frac{\tilde{x}^{\al+1}}{\al+1}\right) \right],
\end{equation}
which, when compared with \eqref{asy-sol} for $k=1$, gives the identification
\begin{equation}\label{u-12}
 u_{-1/2,1/2}= i \sqrt{3}\psi .
\end{equation}

To get the Bethe Ansatz equations (BAE) eq.\eqref{u-12} is used, exploiting the  relation
\begin{equation}\label{u-bae}
 u_{-1/2,1/2}=\psi_{-\frac1{2}}\psi'_{\frac1{2}}-\psi_{\frac1{2}}\psi'_{-\frac1{2}}= i \sqrt{3}\psi_0.
\end{equation}
At this point the solutions $\psi_k$ can be expressed in the basis of the $|z|\rightarrow0$ solutions \eqref{psi0} with
the appropriate normalization to capture
the correct behaviour in the conformal limit
\begin{equation}\label{Q-exp}
 \psi_k =Q^+_{k}(E,\tilde{E})\chi^+_{k}+Q^0_{k}(E,\tilde{E})\chi^0_{k}+Q^-_{k}(E,\tilde{E})\chi^+_{k},
\end{equation}
where
\begin{equation}
 Q^{\pm,0}_k( E,\tilde{E} )=Q^{\pm,0}(\om^{-k 3\al} E,\om^{k 3 \al} \tilde{E} ),
\end{equation}
and
\begin{equation}
\left\{
\begin{aligned}
  \chi^+_{k}&\sim \omega^{k(g+1)}x^{-g} \\
  \chi^0_{k}&\sim  x \\
  \chi^-_{k}&\sim \omega^{-k(g+1)} x^{g+2}.
 \end{aligned}\right.
\end{equation}
The functions $Q^{\pm,0}( E,\tilde{E} )$, introduced in (\ref{Q-exp}), are the  massive  off-critical analogue of the CFT-related
functions $Q^{\pm,0}( E)$ \cite{Dorey:2007zx}, \cite{Bazhanov:1998wj}.
Inserting the expression \eqref{Q-exp} into \eqref{u-bae} and
considering, at first, the terms proportional to $x^{-g}$, a 
functional relation containing $Q^+$ and $Q^0$ is obtained
\begin{equation}
\begin{aligned}
i\sqrt{3}Q^{+}_0&= Q^{+}_{-1/2} Q^{0}_{1/2}\om^{-(g+1)/2}-gQ^+_{1/2}Q^0_{-1/2}\om^{(g+1)/2}+ \\
                 &-Q^{+}_{1/2} Q^{0}_{-1/2}\om^{(g+1)/2}+gQ^+_{-1/2}Q^0_{1/2}\om^{-(g+1)/2} \\
                &=(g+1) \left(Q^{+}_{-1/2} Q^{0}_{1/2}\om^{-(g+1)/2}-Q^{+}_{1/2} Q^{0}_{-1/2}\om^{(g+1)/2}\right)
\end{aligned}
\label{EQEQ}
\end{equation}
with the appropriate $\om$ rotation factors.
Equation (\ref{EQEQ}) can be written in terms of the variable
$\theta= (\al+1)/(3 \al) \ln E$ as:
\begin{equation}
i\sqrt{3}Q^{+}(\theta) = (g+1)\left(Q^{+}(\theta+i\frac{\pi}{3})Q^{0}(\theta-i\frac{\pi}{3})\om^{-\frac{g+1}{2}}
                         -\om^{\frac{g+1}{2}}Q^{+}(\theta-i\frac{\pi}{3})Q^{0}(\theta+i\frac{\pi}{3})\right)
\label{QQ+}
\end{equation}
(with $s$ and $g$ kept constant.)

Considering (\ref{QQ+}) evaluated at $\theta=\theta_n +  i   \pi/3$ and $\theta=\theta_n -  i   \pi/3$ with
$Q^{+}(\theta_n)=0$ to get
\begin{equation}
i\sqrt{3}Q^{+}(\theta_n +  i   \pi/3 ) = (g+1)\left( Q^{+}(\theta_n +  i  2 \pi/3  ) Q^{0}(\theta_n) \om^{-\frac{g+1}{2}}\right)
\end{equation}
and
\begin{equation}
i\sqrt{3}Q^{+}(\theta_n -  i   \pi/3 ) = -(g+1)\left( \om^{\frac{g+1}{2}}Q^{+}(\theta_n-  i  \frac{2\pi}{3}) Q^{0}(\theta_n)\right)
\end{equation}
respectively, and taking the ratio, the sought-after Bethe Ansatz 
equations are obtained \cite{Dorey:1999pv}
\begin{equation}
\frac{Q^{+}(\theta_n +  i  2 \pi/3  ) }{Q^{+}(\theta_n -  i  2 \pi/3  )}
\frac{Q^{+}(\theta_n -  i   \pi/3  ) }{Q^{+}(\theta_n +  i   \pi/3  )}
=- \om^{(g+1)}.
\label{BAE-Q+}
\end{equation}

The same method can be used to find BAE for the $Q^-$ functions. In
conclusion: by defining $\theta^{\pm}$ the zeroes relative to
$Q^{\pm}$, the Bethe Ansatz equations for the two spectral
functions are:
\begin{equation}
\frac{Q^{\pm}(\theta^{\pm}_n +  i  2 \pi/3  ) }{Q^{\pm}(\theta^{\pm}_n -  i  2 \pi/3  )}
\frac{Q^{\pm}(\theta^{\pm}_n -  i   \pi/3  ) }{Q^{\pm}(\theta^{\pm}_n +  i   \pi/3  )}
=- \om^{\pm(g+1)}.
\label{BAE}
\end{equation}
The Bethe Ansatz equations (\ref{BAE}) can be transformed into the
nonlinear integral equation given in section 6 of \cite{Dorey:1999pv} for 
the ground state energy  of the  Izergin-Korepin model or, equivalently, of the  scaling
$q$-state Potts  and tricritical Potts models on a cylinder
geometry. 
This family of systems plays a key role in the study of integrable
quantum field theories in 1+1 dimensions and
statistical mechanical models in 2 dimensions, and
is related to the $\phi_{1,2}$, $\phi_{2,1}$  and 
$\phi_{1,5}$ integrable deformations of minimal  conformal field
theories.
The Ising model in external magnetic field, with its $E_8$-related
mass spectrum \cite{Zamolodchikov:1989fp}, is perhaps the most
physically-interesting  system in the family. A schematic 
dictionary  of the correspondence between the classical objects
introduced here and the `quantum field theory world' is given 
in Table \ref{tab1}.
\begin{table}[h]
\centering
\begin{tabular}{ccc}
\hline\hline
classical modified Bullough-Dodd  &  &     QFT on a cylinder   \\
 \hline
$\theta$               & $\Leftrightarrow$ & particle rapidity  \\
$s^{\al+1}$                      & $\Leftrightarrow$ &   $Rm$ ($R$: circumference, $m$: mass gap)  \\
$g$                        & $\Leftrightarrow$ &  twist parameter \\
$\al$                    & $\Leftrightarrow$ & `$q$' in the
$q$-state Potts model  \\  
\hline\hline
\end{tabular}
\caption{ \label{tab1} \footnotesize The dictionary}
\end{table}
\section{Conclusions}
\label{Conclusions}

The main objective  of this article was to discuss  a particular generalization of the so-called ODE/IM
correspondence to the massive case. The steps taken
by Lukyanov and Zamolodchikov in \cite{Lukyanov:2010rn} clarified
how massive generalisations of the ODE/IM correspondence should be built. 
The results of \cite{Lukyanov:2010rn} concern the relation between
the classical and the quantum versions of the  sinh-Gordon model.
Adapting their discussion it has been possible, here, to relate the
classical (Titzeica-) Bullough-Dodd field equation to the 
Izergin-Korepin massive quantum field theory. The discussion
parallels precisely the conformal case treated in  \cite{Dorey:1999pv}
and links a certain linear problem to a Bethe Ansatz system associated
with the $a_2$ algebra. 
The Bethe Ansatz equations (\ref{BAE}) and the demonstration of a
link with  the
conformal field theory limit case of \cite{Dorey:1999pv}, described in
section \ref{conflim-subsec},  are the main  
results of the paper.

After the first discovery of this type of correspondence, it was
striking how certain functional relations, typically
emerging in the QFT domain, encoded spectral data. In the work of
Lukyanov and Zamolodchikov,
this rich structure was enlarged to contain also the relation
between certain nonlinear partial differential equations and linear
spectral problems, thereby giving further useful insights into the
general structure of the theory. Here,  further support of the
validity of this scheme has been given, enlarging the number
of working cases of the correspondence. 
The Bullough-Dodd equation has been chosen since, with the
sinh-Gordon equation, it is  simplest  representative of the
affine Toda field theories. At this
stage it is fairly clear how a more
general correspondence scheme could be developed starting from  Toda
field theories based on more general Lie algebras. The general scheme
is as follows, where the arrows have been annotated with
labels of sections in this paper
and a reference, to indicate where the corresponding step is further
discussed for the particular
case of the Bullough-Dodd equation:

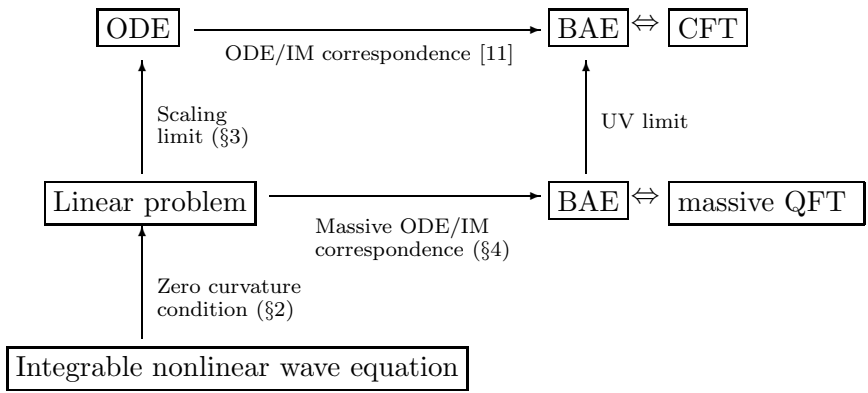
\vspace{.5cm}
\begin{figure}[h]
 \centering
\setlength{\unitlength}{1cm}
\begin{picture}(10,4.5)(0,0)
\put(-1.0,0.0){\framebox{Integrable nonlinear wave equation}}
\put(0.8,0.5){\vector(0,1){1.5}}
\put(1.0,1.1){\scriptsize{Zero curvature}}
\put(1.0,0.8){\scriptsize{condition
(\secref{BDdef-sec})
}}
\put(-0.5,2.2){\framebox{Linear problem}}
\put(0.8,2.7){\vector(0,1){1.5}}
\put(1.0,3.4){\scriptsize{Scaling}}
\put(1.0,3.1){\scriptsize{limit (\secref{conflim-subsec})}}
\put(0.2,4.5){\framebox{ODE}}
\put(1.5,4.6){\vector(1,0){4.6}}
\put(1.9,4.2){\scriptsize{ODE/IM correspondence \cite{Dorey:1999pv}}} 
\put(6.2,4.5){\framebox{BAE}}
\put(7.3,4.6){$\Leftrightarrow$}
\put(7.8,4.5){\framebox{CFT}}
\put(6.2,2.2){\framebox{BAE}}
\put(7.3,2.3){$\Leftrightarrow$}
\put(7.8,2.2){\framebox{massive QFT }}
\put(2.5,2.4){\vector(1,0){3.6}}
\put(3.1,1.9){\scriptsize{Massive ODE/IM }}
\put(3.1,1.6){\scriptsize{correspondence (\secref{BAE-sec}) }}
\put(6.7,2.7){\vector(0,1){1.5}}
\put(6.9,3.3){\scriptsize{UV limit}}
\end{picture}
\end{figure}

Another direction for future work is the generalization of the
correspondence between the Bethe Ansatz and classical integrable
systems to non relativistically-invariant models such as the KdV
equation and its hierarchy, and the generalization from integrable
field theories to integrable lattice models.

In conclusion, the connection between these two, originally
disconnected, domains of mathematics and theoretical physics
gives a hint of a
bigger scheme in the wide framework of classical and quantum integrability.

\bigskip

\noindent{\bf Acknowledgements --}

This project was  partially supported by an INFN grant PI11, an STFC rolling grant  ST/G000433/1
and  by the  Italian MIUR-PRIN contract 2009KHZKRX-007 ``Symmetries of the Universe and of
the Fundamental Interactions''.

%

%
%


\begin{thebibliography}{99}
\raggedright
\parskip 1pt


%
\bibitem{Dorey:1998pt}
P.~Dorey and R.~Tateo, `Anharmonic oscillators, the thermodynamic
Bethe ansatz, and nonlinear integral equations', J.\ Phys.\ A {\bf
32}, L419 (1999) [arXiv:hep-th/9812211].
%
\bibitem{Sha}
Y.~Sibuya, `Global theory of a second-order linear ordinary
differential equation with polynomial coefficient', Amsterdam:
North-Holland, 1975.
%
\bibitem{Voros}
A.~Voros, `Semi-classical correspondence and exact results: the case
of the spectra of homogeneous Schr\"odinger operators', J. Physique
Lett. {\bf 43} (1982) L1.
%
\bibitem{Bazhanov:1994ft}
V.V.~Bazhanov, S.L.~Lukyanov and A.B.~Zamolodchikov, `Integrable
structure of conformal field theory, quantum KdV theory and
thermodynamic Bethe ansatz', Commun.\ Math.\ Phys.\  {\bf 177} (1996) 
381 [hep-th/9412229].
%
\bibitem{Bazhanov:1996dr}
V.V.~Bazhanov, S.L.~Lukyanov and A.B.~Zamolodchikov, `Integrable
Structure of Conformal Field Theory II. Q-operator and DDV
equation', Commun.\ Math.\ Phys.\  {\bf 190} (1997) 247
[arXiv:hep-th/9604044].
%
\bibitem{Bazhanov:1998dq}
V.V.~Bazhanov, S.L.~Lukyanov and A.B.~Zamolodchikov,
`Integrable structure of conformal field theory. III: The Yang-Baxter relation',
Commun.\ Math.\ Phys.\  {\bf 200} (1999) 297
[arXiv:hep-th/9805008].
%
\bibitem{Bax}
R.J.~Baxter,
{`Eight-vertex model in lattice statistics and one-dimensional
anisotropic Heisenberg chain;
1.\,Some fundamental eigenvectors'}, Ann.\ Phys.\ {\bf 76} (1973)
1\toline{24}\,;~
{`2.\,Equivalence to a generalized ice-type model'}, Ann.\ Phys.\ {\bf
76}
(1973) 25\toline{47}\,;~
{`3.\,Eigenvectors of the transfer matrix and Hamiltonian'}, Ann.\
Phys.\ {\bf 76} (1973)
48\toline{71}.
%
\bibitem{Bazhanov:1998wj}
V.V.~Bazhanov, S.L.~Lukyanov and A.B.~Zamolodchikov, `Spectral
determinants for Schroedinger equation and Q-operators of conformal
field theory', J.\ Statist.\ Phys.\  {\bf 102} (2001) 567
[arXiv:hep-th/9812247].
%
%
\bibitem{Suzuki:1999rj}
J.~Suzuki, `Anharmonic oscillators, spectral determinant and short
exact sequence  of $U_q(\hat{sl}(2))$', J.\ Phys.\ A {\bf 32} (1999)
L183 [arXiv:hep-th/9902053].
%
%
\bibitem{Suzuki:1999hu}
J.~Suzuki, `Functional relations in Stokes multipliers and solvable
models related  to $U_q(A_n^{(1)}$)', J.\ Phys.\ A {\bf 33} (2000) 3507
[arXiv:hep-th/9910215].
%
\bibitem{Dorey:1999pv}
P.~Dorey and R.~Tateo,
`Differential equations and integrable models: The SU(3) case',
Nucl.\ Phys.\  B {\bf 571} (2000) 583
[Erratum-ibid.\  B {\bf 603} (2001) 582]
[arXiv:hep-th/9910102].
%
\bibitem{Dorey:2006an} 
P.~Dorey, C.~Dunning, D.~Masoero, J.~Suzuki and R.~Tateo,
`Pseudo-differential equations, and the Bethe ansatz for the classical Lie algebras',
Nucl.\ Phys.\ B {\bf 772} (2007) 249
[arXiv:hep-th/0612298].
%
\bibitem{Dorey:2007zx} 
P.~Dorey, C.~Dunning and R.~Tateo,
`The ODE/IM Correspondence',
J.\ Phys.\ A  {\bf 40} (2007) R205 
[arXiv:hep-th/0703066].
%
\bibitem{Dorey:2009xa} 
P.~Dorey, C.~Dunning and R.~Tateo,
`From PT-symmetric quantum mechanics to conformal field theory',
Pramana {\bf 73}, 217 (2009)
[arXiv:0906.1130 [hep-th]].
%
\bibitem{BB}
C.M.~Bender and S.~Boettcher, `Real spectra in non-hermitian
Hamiltonians having $\cal{PT}$ symmetry', Phys. Rev. Lett. {\bf 80}
(1998) 5243
\physics{9712001}.
%
%
\bibitem{BBN}
C.M.~Bender, S.~Boettcher and P.N.~Meisinger, `$\cal{PT}$ symmetric
quantum mechanics', J. Math. Phys. {\bf 40} (1999) 2201
\quantph{9809072}.
\bibitem{Dorey:1999uk}
P.~Dorey and R.~Tateo,
`On the relation between Stokes multipliers and the T-Q systems of conformal field theory',
Nucl.\ Phys.\  B {\bf 563} (1999) 573
[Erratum-ibid.\  B {\bf 603} (2001) 581]
[arXiv:hep-th/9906219].
%
\bibitem{Dorey:2001uw}
P.~Dorey, C.~Dunning and R.~Tateo,
`Spectral equivalences, Bethe Ansatz equations, and reality
  properties in PT-symmetric quantum mechanics',
  J.\ Phys.\ A  {\bf 34} (2001) 5679
  [arXiv:hep-th/0103051].
%
\bibitem{Shin:2002vu}
  K.C.~~~Shin,
  `On the reality of the eigenvalues for a class of PT-symmetric
  oscillators',
  Commun.\ Math.\ Phys.\  {\bf 229} (2002) 543 
  [arXiv:math-ph/0201013].
%
\bibitem{Dorey:2009tc}
  P.~Dorey, C.~Dunning, A.~Lishman and R.~Tateo,
  `PT symmetry breaking and exceptional points for a class of
  inhomogeneous
  complex potentials',
  J.\ Phys.\ A  {\bf 42} (2009) 465302
  [arXiv:0907.3673 [hep-th]].
%
\bibitem{Bazhanov:2003ua}
  V.V.~Bazhanov, S.L.~Lukyanov and A.M.~Tsvelik,
  `Analytical results for the Coqblin-Schrieffer model with
  generalized magnetic fields',
  Phys.\ Rev.\ B {\bf 68} (2003) 094427
  [arXiv:cond-mat/0305237].
%
\bibitem{Gritsev:2006}
V.~Gritsev, E.~Altman, E.~Demler and A.~Polkovnikov, 
`Full quantum
distribution of contrast in interference experiments between
interacting one dimensional Bose liquids', 
Nature Physics {\bf 2} (2006) 705
[arXiv:cond-mat/0602475].
%
\bibitem{Lamacraft:2008}
 A.~Lamacraft and P.~Fendley,
 `Order parameter statistics in the critical quantum Ising chain',
 Phys.\ Rev.\ Lett.\ {\bf 100} (2008) 165706
 [arXiv:0802.1246 [cond-mat]].
%
\bibitem{Gaiotto:2009hg}
D.~Gaiotto, G.W.~Moore and A.~Neitzke,
`Wall-crossing, Hitchin Systems, and the WKB Approximation'
[arXiv:0907.3987 [hep-th]].
%
\bibitem{Alday:2009dv}
L.F.~Alday, D.~Gaiotto and J.~Maldacena,
`Thermodynamic Bubble Ansatz',
JHEP {\bf 1109} (2011) 032
[arXiv:0911.4708 [hep-th]].
%
\bibitem{Lukyanov:2010rn}
S.L.~Lukyanov and A.~B.~Zamolodchikov,
`Quantum Sine(h)-Gordon Model and Classical Integrable Equations',
JHEP {\bf 1007} (2010) 008
[arXiv:1003.5333 [math-ph]].
%
\bibitem{Dodd:1977bi}
R.K.~Dodd and R.K.~Bullough,
`Polynomial Conserved Densities for the Sine-Gordon Equations',
Proc.\ Roy.\ Soc.\ Lond.\  {\bf A352} (1977) 481.
%
\bibitem{Tzitzeica}
G.~Tzitzeica,
`Sur une Nouvelle Classe de Surfaces',
C.\ R.\ Acad.\ Sci.\ Paris, {\bf 144} (1907) 1257.
%
\bibitem{IzKo}
A.G. Izergin and V.E. Korepin, ‘The inverse scattering method approach to the
quantum Shabat-Mikhailov model’, Comm. Math. Phys. 79 (1981) 303–331
%
\bibitem{ZamInt}
A.B. Zamolodchikov, ‘Integrable Field Theory from Conformal Field Theory’,
Proceedings of Taniguchi Symposium, Kyoto (1988)
%
\bibitem{Smirnov:1991uw} 
F.A.~Smirnov,
`Exact S matrices for phi(1,2) perturbated minimal models of conformal field theory',
Int.\ J.\ Mod.\ Phys.\ A {\bf 6}, 1407 (1991).
%
\bibitem{Potts}
R.B.~Potts, `Some generalized order-disorder transformations',  Proc.
Cambridge Phil. Soc. 48 (1952) 106.
%
\bibitem{Warnaar:1992gj} 
S.O.~Warnaar, B.~Nienhuis and K.A.~Seaton,
`New construction of solvable lattice models including an Ising model
in a field',
Phys.\ Rev.\ Lett.\  {\bf 69}, 710 (1992).
%
\bibitem{TesiFaldella}
S.~Faldella, `Correspondence between Classical and Quantum Integrability', 
(Master Thesis in ``Physics  of Fundamental  Interactions'' July 2011,
University of Turin, Italy).
%
\bibitem{Toda}
A.V. Mikha{\u{\i}}lov,
`Integrability of a two-dimensional generalization of the Toda chain',
JETP Letters, 30(7) (1979) 414.
%
%
\bibitem{Ablowitz}
M.~Ablowitz and P.~Clarkson, `Solitons, Nonlinear Evolution Equations and
Inverse Scattering', Cambridge University Press, 1991.
%
\bibitem{Zamolodchikov:1989fp}
A.B.~Zamolodchikov,
`Integrals of Motion and S Matrix of the (Scaled) T=T(c) Ising Model
with Magnetic Field',
Int.\ J.\ Mod.\ Phys.\  A {\bf 4} (1989) 4235.
%
\end{thebibliography}
\end{document}